\documentclass[fleqn,10pt]{SelfArx} 

\definecolor{color1}{RGB}{0,0,90} 
\definecolor{color2}{RGB}{0,20,20} 

\usepackage{hyperref} 
\hypersetup{hidelinks,colorlinks,breaklinks=true,urlcolor=color2,citecolor=color1,linkcolor=color1,bookmarksopen=false,pdftitle={Title},pdfauthor={Author},urlcolor=blue}

\usepackage{graphicx}
\usepackage[square]{natbib}
\usepackage{amsmath}
\usepackage{multirow}
\usepackage{subfigure}
\usepackage{siunitx}
\usepackage{geometry}
\usepackage{pdflscape}

\newcommand{\n}[1]{\mathrm{#1}}

\JournalInfo{Published in Energy Conversion and Management, Vol. 156, 264-268, 2018} 
\Archive{\href{http://dx.doi.org/10.1016/j.enconman.2017.11.009}{DOI: 10.1016/j.enconman.2017.11.009}} 

\PaperTitle{The maximum theoretical performance of unconcentrated solar photovoltaic and thermoelectric generator systems} 

\Authors{R. Bj\o{}rk, K. K. Nielsen} 
\affiliation{\textit{Department of Energy Conversion and Storage, Technical University of Denmark - DTU, Frederiksborgvej 399, DK-4000 Roskilde, Denmark}} 
\affiliation{*\textbf{Corresponding author}: rabj@dtu.dk} 

\newcommand{\keywordname}{Keywords} 

\Abstract{The maximum efficiency for photovoltaic (PV) and thermoelectric generator (TEG) systems without concentration is investigated. Both a combined system where the TEG is mounted directly on the back of the PV and a tandem system where the incoming sunlight is split, and the short wavelength radiation is sent to the PV and the long wavelength to the TEG, are considered. An analytical model based on the Shockley-Queisser efficiency limit for PVs and the TEG figure of merit parameter $zT$ is presented. It is shown that for non-concentrated sunlight, even if the TEG operates at the Carnot efficiency and the PV performance is assumed independent of temperature, the maximum increase in efficiency is 4.5 percentage points (pp.) for the combined case and 1.8 pp. for the tandem case compared to a stand alone PV. For a more realistic case with a temperature dependent PV and a realistic TEG, the gain in performance is much lower. For the combined PV and TEG system it is shown that a minimum $zT$ value is needed in order for the system to be more efficient than a stand alone PV system.}

\begin{document}

\flushbottom 

\maketitle 


\thispagestyle{empty} 

\section{Introduction}
Conversion of solar radiation directly to electricity with as high efficiency as possible is of immense interest to society. Photovoltaic devices (PV) that can directly convert parts of the solar spectrum to radiation receive a significant amount of attention, but recently additional energy conversion technologies that can utilize the remaining part of the solar spectrum have come into focus. The part of the solar spectrum not converted by a PV is typically turned into heat, and this has lead to an increased focus on coupling a PV with a thermoelectric generator (TEG), which can convert a flow of heat directly to electrical energy through the Seebeck effect.

\begin{figure*}[!t]
  \centering
  \includegraphics[width=2\columnwidth]{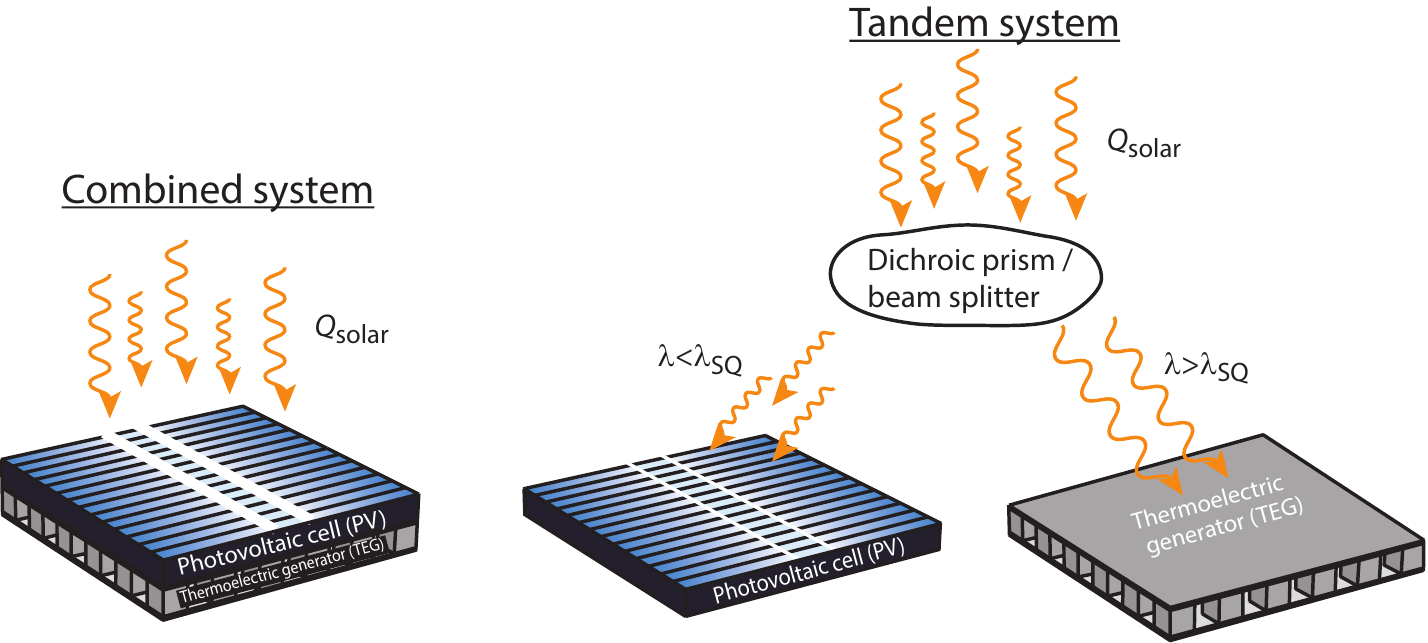}
  \caption{An illustration of the two systems considered. In the tandem system, the incoming solar radiation, $Q_\n{solar}$, is split between the PV and TEG at a wavelength $\lambda_{SQ}$.}
  \label{Fig_Illustration_PV_TEG_II}
\end{figure*}

A combination of a PV and a TEG could potentially have a higher efficiency, i.e. be able to convert a larger fraction of the incoming solar radiation into electricity, than a PV alone. There are two ways to realize a system containing both a PV and a TEG device. In a combined system, the TEG is mounted directly on the back of the PV. The heat absorbed by the PV is transferred through the TEG, generating electricity. In a tandem system, the incoming sunlight is split in wavelength by a wavelength separating device (beam splitter/dichroic prism), and the short wavelength radiation is sent to the PV while the long wavelength goes to the TEG. The two kinds of PV and TEG systems are illustrated in Fig. \ref{Fig_Illustration_PV_TEG_II}.

The tandem system has been studied previously \cite{Luque_1999,Zhang_2005,Kraemer_2008}, and experimental systems have also been realized, but only with small gains in open circuit voltage and efficiency \cite{Mizoshiri_2012,Ju_2012}. A review of the field from an experimental point of view was presented in \citet{Sundarraj_2014}. The combined system has also been considered in some detail \cite{Sark_2011,Kiflemariam_2014,Liao_2014,Zhang_2014,Attivissimo_2015,Wu_2015}, with a number of studies focussing on the effect of concentration on the performance of the combined system. Recently, Beeri et al. investigated an experimental setup with a PV and a Bi$_2$Te$_3$ TEG and found that a concentration factor of 200 was needed before the TEG electrical contribution started to dominate \cite{Beeri_2015}. However, the overall system efficiency did not increase as the concentration was increased. In another experimental study Kossyvakis et al. examined a PV and TEG system, where the PV was either poly-Si or dye-sensitized based. The authors observed that only for the lower efficiency dye-sensitized cell was the resulting system performance similar to the operation of the PV alone \cite{Kossyvakis_2016}. Finally, in a numerical study Lamba et al. modelled a combined PV and TEG system and found that increasing the concentration from one to five decreased the total system efficiency, due to the temperature dependency of the PV \cite{Lamba_2016}. In general, the conclusion is that with concentration a gain in performance is possible \cite{Vorobiev_2006,Lin_2014,Xu_2014,Najafi_2013,Fisac_2014}, but without concentration and using commercial TEG modules, no gain in performance is seen \cite{Lorenzi_2015,Bjoerk_2015b}.

In this work, we determine the maximum performance for unconcentrated systems. We present the maximum theoretical performance of both combined and tandem PV and TEG systems for unconcentrated systems. We present an analytical approach that only relies on established material properties, such as the thermoelectric figure of merit, $zT$, and the PV Shockley-Queisser limit. The results are applicable to all current and future PV and TEG technologies and materials, as long as the system is not exposed to concentrated light. A somewhat similar approach was considered by \citet{Lorenzi_2015}, except the performance was not given as a function of $zT$.

\section{The studied systems}
We consider two kinds of PV and TEG systems, namely a combined system and a tandem system. In a combined system, the TEG is mounted directly on the back of the PV. The hot side temperature of the TEG is thus equal to the temperature of the PV. In a tandem system, the incoming sunlight is split in wavelength by a dichroic beam splitter at a wavelength $\lambda_{SQ}$, such that the low energy (long wavelength) light is sent to the TEG, while the high energy (short wavelength) light is sent to the PV.

\subsection{TEG properties}
We consider a general TEG, for which the efficiency is given according to the figure of merit, $zT$, of the device. The efficiency as a function of $zT$ is given by \citep{TE_Handbook_Ch9}
\begin{eqnarray}
\eta_\n{TEG} = \frac{T_\n{H}-T_\n{C}}{T_\n{H}}\frac{\sqrt{1+zT}}{\sqrt{1+zT}+\frac{T_\n{C}}{T_\n{H}}}\label{eq_eff_TEG}
\end{eqnarray}
where $T_\n{H}$ and $T_\n{C}$ are the hot and cold side temperatures, respectively, and the figure of merit is
\begin{eqnarray}
zT = \frac{\sigma{}S^2T}{\kappa}
\end{eqnarray}
where $\sigma$ is the electrical conductivity, $S$ is the Seebeck coefficient, $T$ is the temperature and $\kappa$ is the thermal conductivity. Here we take $zT$ to be a material parameter, as is traditionally done in the field of thermoelectrics. For current thermoelectric materials $zT$ varies between 0.5 to 1.5, and increases significantly with temperature \cite{Ngan_2014,Bjoerk_2015}. It is noted that a TEG operates between a hot, $T_\mathrm{H}$, and a cold, $T_\mathrm{C}$, temperature and that $zT$ is a function of temperature. Thus $zT$ in Eq. (\ref{eq_eff_TEG}) can be considered as the average value over the temperature range $T_\mathrm{C}$ to $T_\mathrm{H}$. A TEG can be designed with an arbitrary thickness or area, and thus its thermal resistance can be completely controlled without affecting its $zT$ value.

We also consider the maximum theoretical efficiency possible when operating a heat engine between two thermal reservoirs, namely a Carnot engine. For this the efficiency is given as
\begin{eqnarray}
\eta_\n{Carnot} =1- \frac{T_\n{C}}{T_\n{H}}~.
\end{eqnarray}
This corresponds to a TEG with $zT\rightarrow{}\infty{}$.

\subsection{PV properties}
We consider a single p-n junction PV that performs at the maximum theoretical efficiency, known as the Shockley-Queisser (SQ) limit \cite{Shockley_1961}. In this limit losses occurring due to impedance matching, recombination, blackbody radiation and spectrum losses are all accounted for. The efficiency of the SQ PV depends on the band gap of the p-n junction, or correspondingly on the wavelength, here termed $\lambda_\n{SQ}$. The efficiency has been tabulated as a function of $\lambda_\n{SQ}$ for a PV cell operating at 25 $^\circ$C irradiated by the AM 1.5G spectral irradiance (ASTM G173-03) and considering radiative emission only from the front side due to a perfect reflector at the rear side \cite{Ruhle_2016}. These tabulated values are used throughout this work.

For the case of the combined PV and TEG, the PV will increase in temperature. The degradation in PV efficiency with temperature is given by
\begin{eqnarray}
\eta_\n{PV} = \eta_{T_\n{ref}}\left(1+\beta(T_\n{ref}-T)\right)\label{eq_eff_PV}
\end{eqnarray}
where $\eta_{T_\n{ref}}$ is the efficiency of the PV at the reference temperature $T_\n{ref}$, taken to be 25 $^\circ$C, and $\beta$ is the temperature coefficient \cite{Skoplaki_2009}. Here $\eta_{T_\n{ref}}$ is given by the Shockley-Queisser (SQ) limit \cite{Ruhle_2016}. For reference, the values for the temperature coefficient for typical commercial PVs are 0.392, 0.110, 0.353 and 0.205 \% K$^{-1}$ for crystalline Si (c-Si) \cite{Skoplaki_2009}, amorphous Si (a-Si) \cite{Skoplaki_2009}, copper indium gallium (di)selenide (CIGS) \cite{TSMC_2015,Hulk_2015,Solibro_2015} and cadmium telluride (CdTe) \cite{Singh_2012} PVs, respectively.

\section{The physical models}\label{Sec_model}
We assume an incoming solar flux of 1000.37 Wm$^{-2}$ and the full AM 1.5G spectral irradiance (ASTM G173-03), with an intensity denoted by $I_\n{AM1.5}$ as a function of frequency. First, we consider the combined PV and TEG system. Here the incoming solar flux will either be converted to electricity by the PV or to heat, raising the temperature of the PV and the hot side of the TEG.  The efficiency of the PV is given by the SQ limit as a function of wavelength as described above.

The amount of heat available to the TEG depends on the efficiency of the PV through $Q_\n{hot} = Q_\n{sun}(1-\eta_\n{PV})$ plus an additional heat $Q_\n{amb}$, which is the radiation from the ambient. This accounts for the fact that the PV surface can never drop to a temperature below the ambient temperature. For the combined case, the incoming radiation is simply given as the integral over the full AM 1.5G spectrum, equalling $Q_\n{sun}=1000.37$ Wm$^{-2}$. In equilibrium the available heat $Q_\n{hot}+Q_\n{amb}$ must be equal to the heat radiated from the hot surfaces of the PV and TEG, $Q_\n{rad}$, and the heat conducted through the TEG to its cold side, $Q_\n{TEG}$, i.e.
\begin{eqnarray}\label{eq_Qhot}
Q_\n{hot} + Q_\n{amb}= Q_\n{TEG} + Q_\n{rad}.
\end{eqnarray}
Here we have ignored all other losses than radiative. Convective losses could also be included as a loss term in Eq. \ref{eq_Qhot}, but this would reduce the efficiency further. As we are interested in the ideal performance convective losses are ignored.

The heat radiated from the hot surfaces of the PV and TEG is given by Stefan-Boltzmanns law, as is the incoming ambient radiation. We thus have for the hot side temperature, $T$,
\begin{eqnarray}\label{eq_Qall}
\sigma_\n{SB}T^4 = Q_\n{sun}\left(1-\eta_\n{PV}(T,\lambda_\n{SQ})\right) + \sigma_\n{SB}T_\n{amb}^4 - Q_\n{TEG};
\end{eqnarray}
Here $\eta_\n{PV}$ is a function of temperature and $\lambda_\n{SQ}$, and thus the above equation cannot be solved analytically. However, a solution can be obtained for values of $Q_\n{TEG}$ by iteration, i.e. by guessing a temperature and solving the terms in Eq. \ref{eq_Qall} until the equation is satisfied. The ambient temperature is taken to be $T_\n{amb}=25$ $^\circ$C. The parameter $\sigma_\n{SB}= 5.670367*10^{-8}$ Wm$^{-2}$K$^{-4}$ is the Stefan-Boltzmann constant.
We assume an emissivity of 1 in Stefan-Boltzmanns law, in order for the PV (and TEG) to absorb as much sunlight as possible. Note that the emissivity is not considered to be a function of wavelength, as is discussed below.

For the tandem system, the model is simpler. Here the heat available to the TEG is the fraction of the solar spectrum above the chosen $\lambda_\n{SQ}$, i.e.
\begin{eqnarray}
Q_\n{hot} = \int_{\lambda_\n{SQ}}^{\infty}I_\n{AM1.5}(\lambda)d\lambda.
\end{eqnarray}
The efficiency of the PV in this system does not dependent on temperature, as the PV is assumed to be cooled to the ambient temperature. The efficiency of the PV is thus given by the chosen $\lambda_\n{SQ}$ according to the SQ limit. For the TEG, there is a balance between the available heat, the heat transferred through the TEG and the heat radiated to the surroundings. Thus Eq. \ref{eq_Qhot} still holds except with $Q_\n{hot}$ as given above. Contrary to the combined case, for the tandem case this equation can be solved analytically for the temperature of the TEG, for any value of $Q_\n{TEG}$, as the PV efficiency does not depend on the temperature of the TEG.

In design of solar collectors and solar thermoelectric generators a useful trick in order to increase the temperature of the system is to use a material for which the emissivity depends on wavelength, i.e. a non black body system. Using such a material the high wavelength radiation can be absorbed and turned to low wavelength heat, where the emissivity has been designed to be lower. This can raise the temperature of the system significantly and thus increase performance of TEG devices. However, here this trick is not considered, as the PV absorbs the high wavelength radiation. Thus here we assume a constant emissivity of 1, that is not a function of wavelength, to obtain as efficient a system as possible.

For both the tandem and the combined case, once the temperature is found as a function of $Q_\n{TEG}$ the maximum total efficiency of the PV and the TEG system is determined, as a function of $Q_\n{TEG}$. This means that the TEG is chosen to have a thickness or area, i.e. a thermal resistance, that maximizes the total efficiency of the PV and TEG system. In this regard it matters not how many legs or what thermoelectric material the TEG is made of. The efficiency of a TEG does not depend on how much heat is conducted through it, and thus the TEG can be chosen with a thickness that maximizes the total efficiency. As stated previously this is done as a function of the thermoelectric figure of merit parameter $zT$, which is the only parameter needed to specify the performance of a TEG.
It should be noted that in both the combined and tandem case, we assume that the cold side of the TEG, and for the tandem case also the PV, can be cooled for free.

\section{Results}
\begin{figure}[!t]
\subfigure[]{\includegraphics[width=0.45\textwidth]{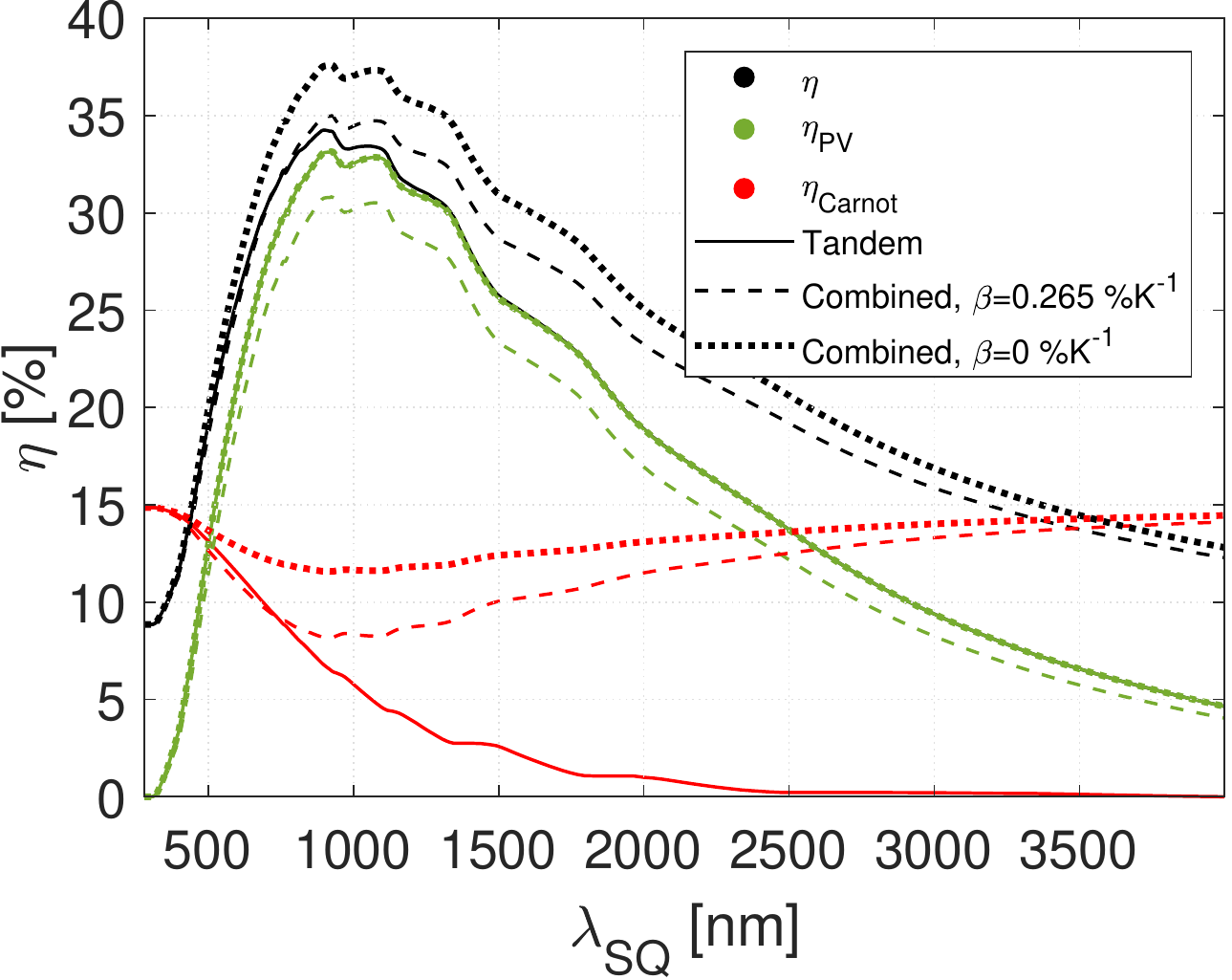}}
\subfigure[]{\includegraphics[width=0.45\textwidth]{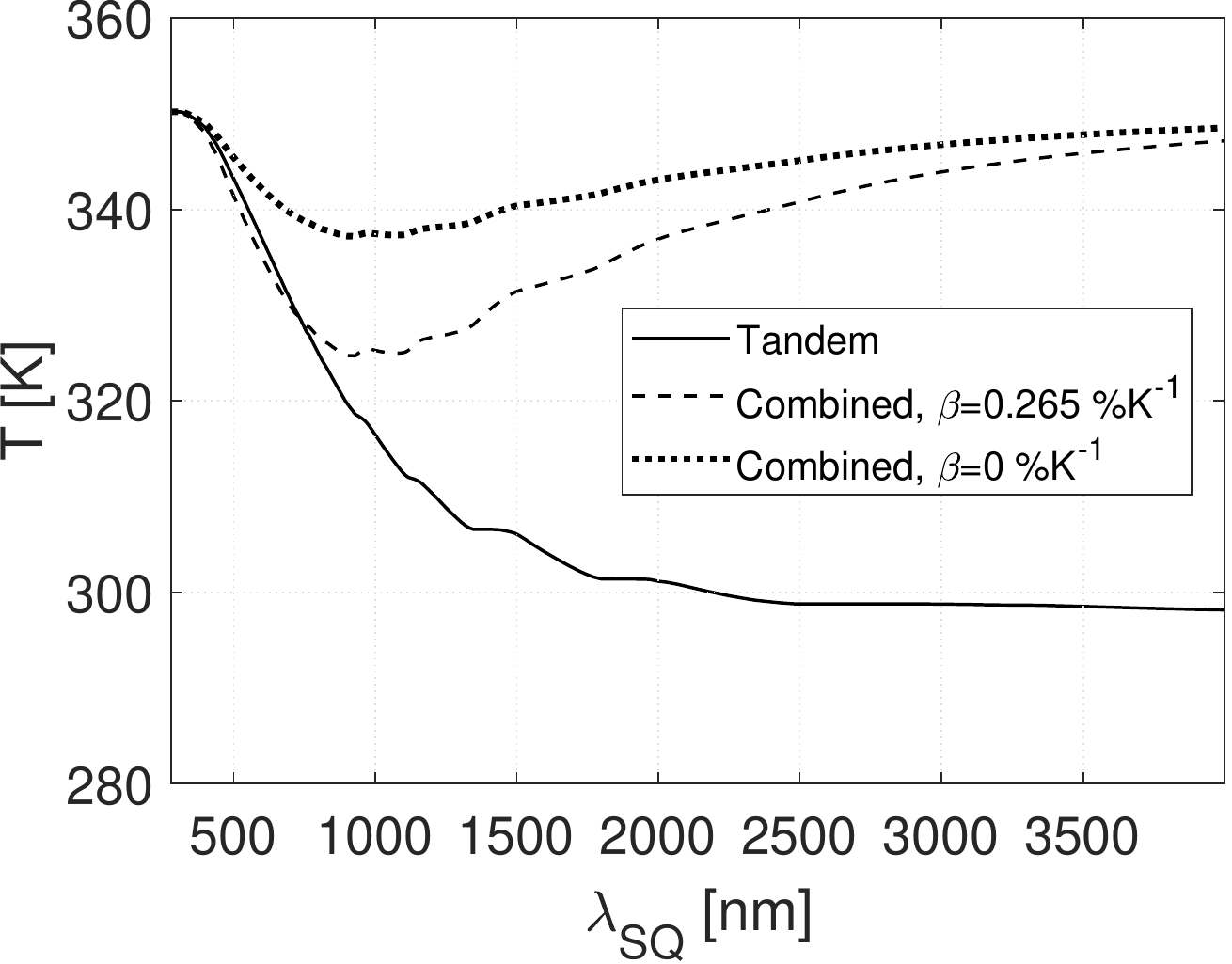}}
\caption{a) The efficiency of a PV and a Carnot engine in tandem and combined configurations as a function of $\lambda_\n{SQ}$. b) The temperature of the hot side of the Carnot engine as a function of $\lambda_\n{SQ}$.}\label{Fig_Eff_SQ_plot}
\end{figure}

We first consider the maximum performance of both a system with ideal components and a system with a current generation high performing PV. For the latter case, this corresponds to a PV with a $\beta$-coefficient of 0.265 \% K$^{-1}$, which is the average $\beta$-coefficient of c-Si, a-si, CIGS and CdTe PV cells. As TEG we initially consider a device with $zT\rightarrow\infty$, i.e. a Carnot engine, to establish the maximum efficiency possible. The total efficiency, i.e. the fraction of incoming sunlight converted to electricity, is shown as a function of $\lambda_\n{SQ}$ in Fig. \ref{Fig_Eff_SQ_plot}a. The efficiency of the PV peaks at a value of $\lambda_\n{SQ}\approx 900$ nm, which is the optimal value for $\lambda_\n{SQ}$ \cite{Ruhle_2016}. The figure breaks the efficiency down into the PV, Carnot and combined efficiencies. As can be seen, adding a Carnot engine will, in the combined case with $\beta=0.265$ \% K$^{-1}$ raise the efficiency by approx. 2 percentage points (pp.) while for the tandem case the gain in efficiency is 1.2 pp., compared to the stand alone PV. For the case of $\beta=0$ \% K$^{-1}$, the gain in efficiency is 4.5 pp. Note that the $\eta_{PV}$ values for the tandem case are the tabulated values from Ref. \cite{Ruhle_2016}, i.e. the maximum stand-alone PV efficiency. Also note that there is no difference between the performance of the PV in the tandem case and in the combined case with $\beta=0$ \% K$^{-1}$.

For each value of $\lambda_\n{SQ}$ the heat flow through the Carnot engine is varied as described in Sec. \ref{Sec_model} to determine the optimal combined efficiency of the PV and TEG. Once this thermal resistance has been found, the temperature is computed. This is shown in Fig. \ref{Fig_Eff_SQ_plot}b. The temperature varies between 350 K and room temperature, with the temperature being highest for the combined case. It is also seen that the combined system with $\beta=0$ \% K$^{-1}$ has maximum efficiency at a higher temperature than the system with $\beta=0.265$ \% K$^{-1}$, where the efficiency of the PV decrease with temperature.

The reason that the gain in efficiency is only a few percentage points is because the temperature of the hot side remains low. The reason for this is that the temperature cannot be raised more for systems without solar concentration. For an unconcentrated black-body system at the surface of the Earth, the maximum temperature that can be reached can be computed trivially from Stefan-Boltzmanns law as $T=\left(\frac{1000.37\ \mathrm{W/m}^2}{\sigma_\n{SB}}\right)^{1/4}=364.5$ K, provided all the incoming radiation is absorbed and reemitted.

\begin{figure}[!b]
  \centering
  \includegraphics[width=1\columnwidth]{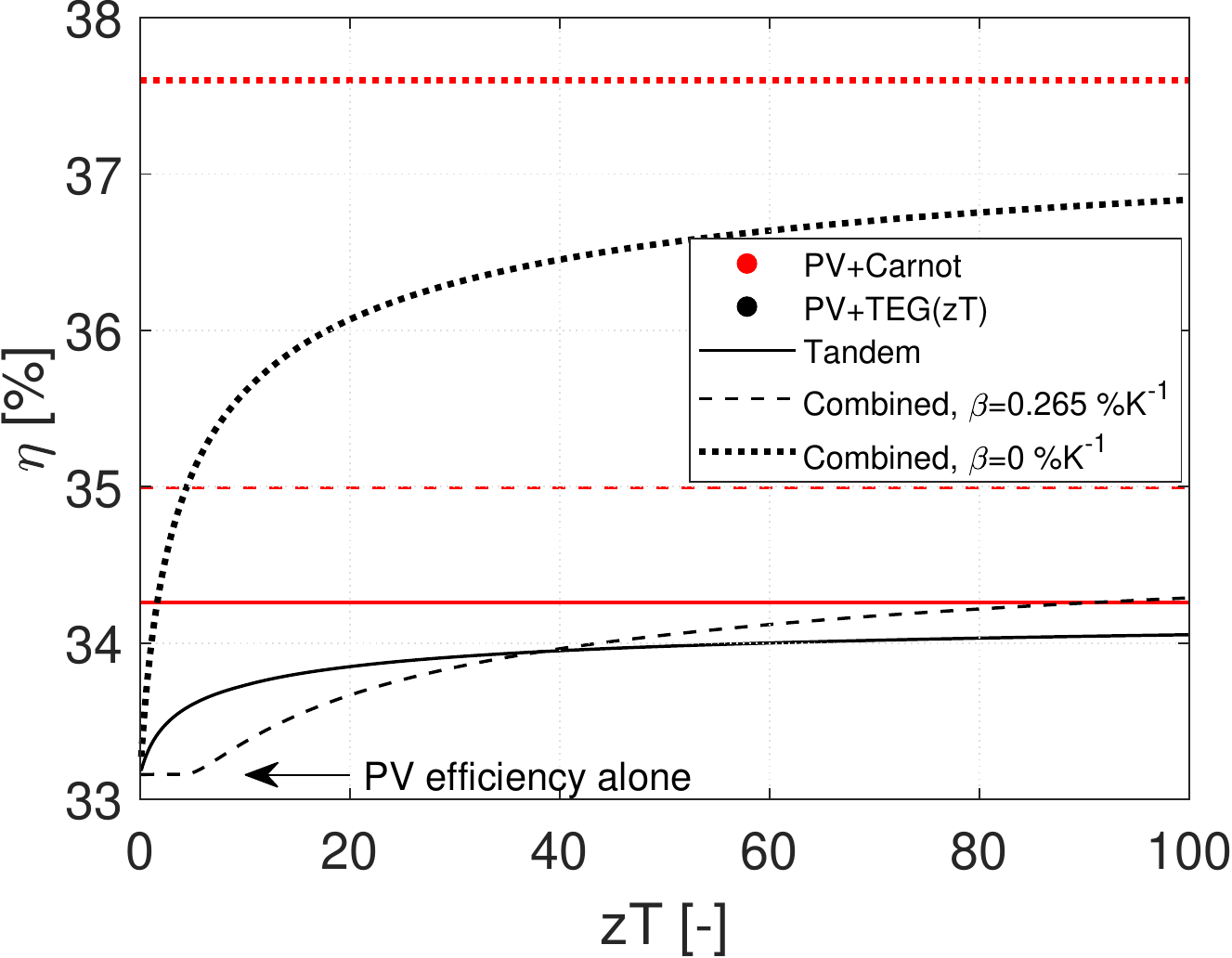}
  \caption{The maximum efficiency of a tandem and a combined PV and TEG system as a function of $zT$.}
  \label{Fig_Eff_zT_plot}
\end{figure}

Instead of the Carnot engine considered above, we now consider a TEG with a given $zT$ value. The computations done are similar to those described for the Carnot engine above, i.e. the optimal thermal resistance is determined for each value of $\lambda_\n{SQ}$. The maximum efficiency is then determined for all values of $\lambda_\n{SQ}$, for a range of $zT$ values. The resulting maximum efficiency as a function of $zT$ is shown in Fig. \ref{Fig_Eff_zT_plot}. Note that the highest performing thermoelectric material today has $zT\approx2$, and this is at a much higher temperature than the maximum 364 K considered here. As can be seen from the figure, for the tandem case a TEG with an arbitrarily low value of $zT$ will result in a gain in efficiency. This is because the incoming light is split between the TEG and the PV, and thus the TEG will always contribute with a net efficiency to the overall system. For the combined case, this is not the case. Here a positive value of $\beta$ means that any temperature above room temperature will result in a decreased performance of the PV. If the $zT$ value of the TEG is too low, the power produced by the TEG is lower than the power lost by heating the PV. If this is the case, the optimal configuration is one where the TEG is simply not present. There is thus a minimum value of $zT$ that the TEG must have in the combined case, in order for the PV and TEG system to be more efficient than the PV alone.

\begin{figure}
  \centering
  \includegraphics[width=1\columnwidth]{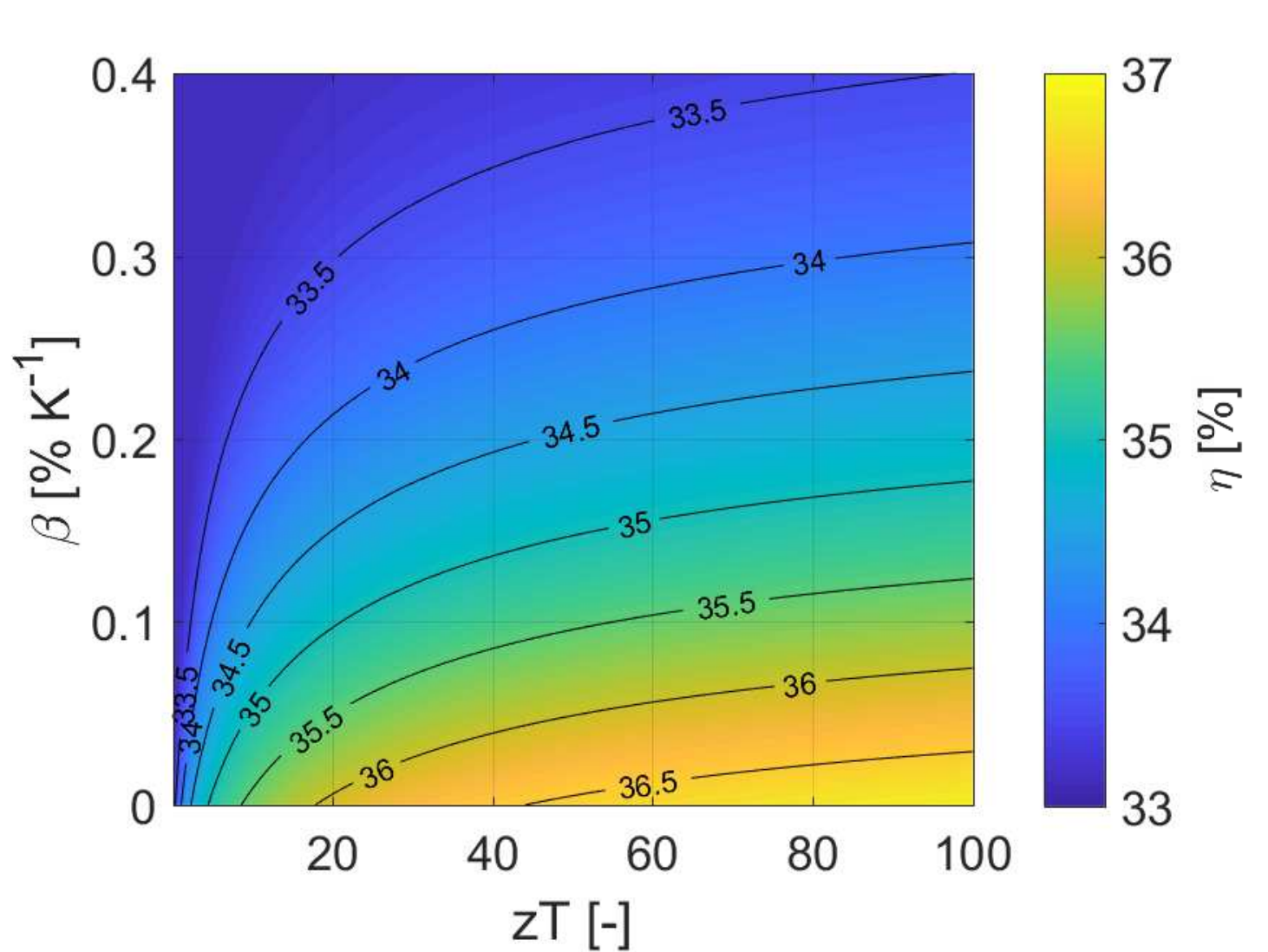}
  \caption{The efficiency as a function of $zT$ and $\beta$ for a combined system. The sole PV efficiency as function of $\beta$ is the value at $zT=0$.}
  \label{Fig_eff_beta_zT_SQ}
\end{figure}

The corresponding maximum efficiency of the combined case as a function of $zT$ and $\beta$ is shown in Fig. \ref{Fig_eff_beta_zT_SQ}. As can be seen the gain in efficiency by adding a TEG to a PV is only a few percentage points, even for the ideal case of $\beta=0$ and an extremely efficient thermoelectric module with very high $zT$ value. This very small gain in performance for an unconcentrated device is completely in line with experiments, which see only a minor gain in output power \cite{Mizoshiri_2012,Ju_2012}, and also in line with previous modeling results \cite{Bjoerk_2015b,Lorenzi_2015}. However, here we have established that this will always be the case, regardless of how efficient TEGs are made in the future. We have thus shown that is it necessary to concentrate the solar radiation \cite{Zhang_2014} in order to gain a high performance of a PV-TEG system.

\begin{figure}[!t]
  \centering
  \includegraphics[width=1\columnwidth]{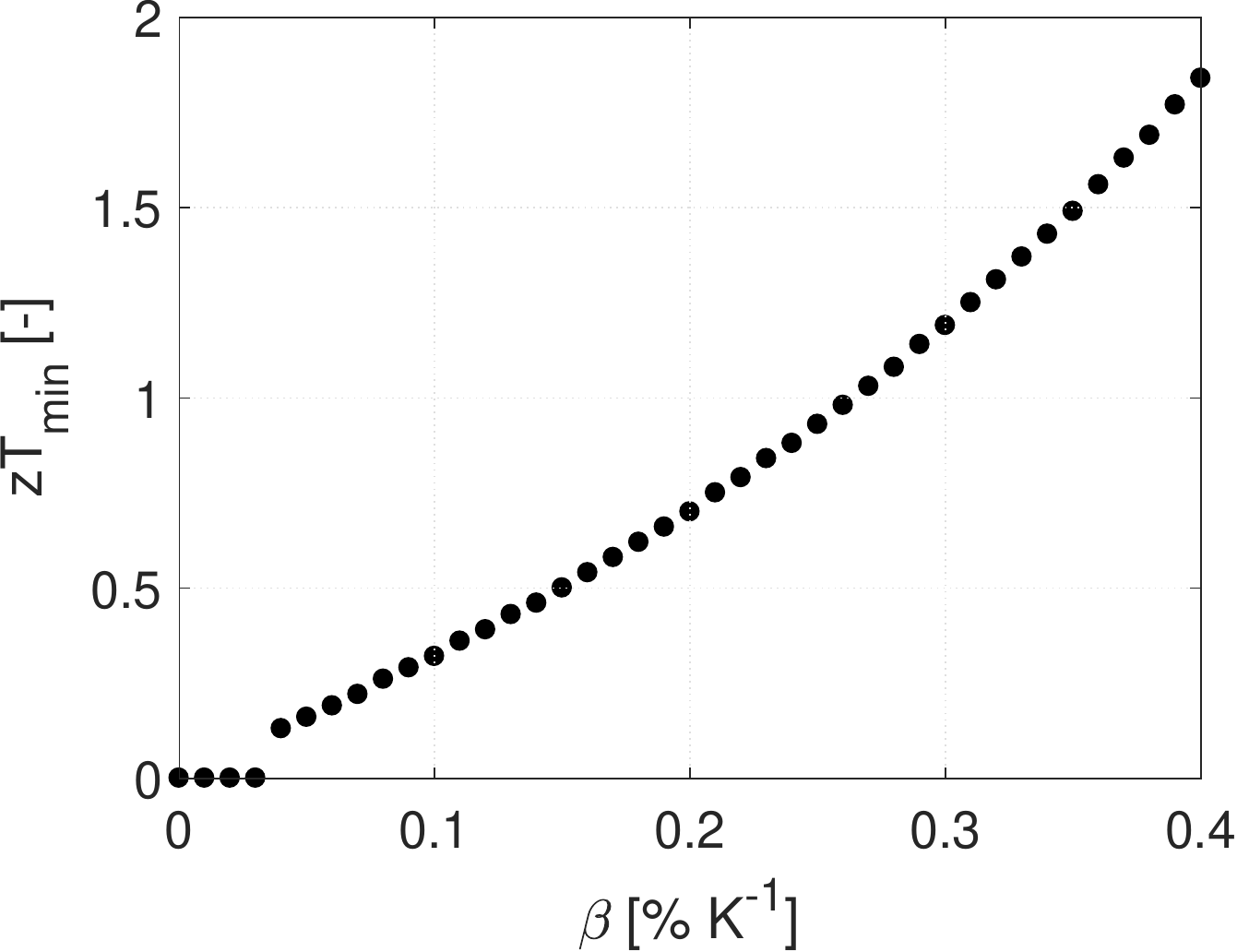}
  \caption{The minimum value of $zT$ needed in a combined system in order to produce an efficiency higher than the stand alone PV efficiency, as a function of $\beta$. The line is a guide to the eye.}
  \label{Fig_zT_min_plot}
\end{figure}

Finally, we saw in the analysis described above that a minimum value of $zT$ was required in order for the combined PV and TEG system to be more efficient than a PV alone. This minimum $zT$ value has been determined as a function of $\beta$ and is shown in Fig. \ref{Fig_zT_min_plot}. As seen from the figure, as the performance of the PV decreases, the less efficient the TEG needs to be in order to contribute positively in a combined system.

\section{Conclusion}
We have investigated unconcentrated PV and TEG systems, both a combined system where the TEG is mounted directly on the back of the PV and a tandem system where the incoming sunlight is split and the short wavelength radiation is sent to the PV and the long wavelength to the TEG. An analytical model was presented that accounts for the incoming radiation from the sun and is based on the Shockley-Queisser efficiency limit for PVs and the efficiency of a TEG as a function of the material figure of merit $zT$. From this model the performance of both the tandem and the combined system was calculated as a function of the PV efficiency, $zT$ and temperature dependence of the PV, $\beta$. It was shown that even if the TEG is as efficient as a Carnot engine and the performance of the PV does not depend on temperature, the maximum increase in efficiency is 4.5 percentage points (pp.) for the combined case and 1.8 pp. for the tandem case. For a more realistic case with a temperature dependent PV and a more realistic TEG, the gain in performance is significantly lower. For the combined PV and TEG system it was also shown that a minimum $zT$ value is needed in order for the system to be more efficient than a stand alone PV system.

\end{document}